\documentstyle[12pt,aasms4,psfig]{article}

\received{1997 July}
\accepted{1997 September}
\singlespace

\begin{document}

\title{MAGNETIC FIELD EFFECTS ON THE HEAD STRUCTURE OF
PROTOSTELLAR JETS}

\author{Adriano H. Cerqueira \footnote{(adriano,
dalpino)@iagusp.usp.br}, Elisabete M. de Gouveia Dal Pino
$^1$} \affil{Instituto Astron\^{o}mico e Geof\'{\i}sico, Universidade
de S\~{a}o Paulo, \\ Av.  Miguel St\'efano 4200, (04301-904) S\~{a}o
Paulo - SP, BRAZIL \\ and}

\author{Marc Herant \footnote {mrc@t6-serv.lanl.gov. Current address:
Washington University School of Medicine, Box 8107, 660 S. Euclid, St.
Louis, MO 63110.}} \affil{Theory Division, MS B228, Los Alamos National
Laboratory, Los Alamos, NM 87545}

\vskip 2.0 cm

\centerline {\bf The Astrophysical Journal Letters, in press}

\begin{abstract}

We present the results of three-dimensional smooth particle
magnetohydrodynamics  numerical simulations of supermagnetosonic,
overdense, radiatively cooling jets. Together with a baseline
non-magnetic calculation, two initial magnetic configurations (in
$\sim$ equipartition with the gas) are considered: (i) a helical and
(ii) a longitudinal field  which permeate both the jet and the ambient
medium. We find that magnetic fields have important effects on the
dynamics and structure of radiative cooling jets, especially at the
head. The presence of a helical field suppresses the formation of the
clumpy structure which is found to develop at the head of $purely$
$hydrodynamical$ jets by fragmentation of the cold shell of shocked
material. On the other hand, a cooling jet embedded in a longitudinal
magnetic field retains clumpy morphology at its head.  This fragmented
structure resembles the knotty pattern commonly observed in HH objects
behind the bow shocks of protostellar jets. This suggests that a strong
(equipartition) helical magnetic field configuration is ruled out at
the jet head.  Therefore, if strong magnetic fields are present, they
are probably predominantly longitudinal in those regions.  In both
magnetic configurations, we find that the confining pressure of the
cocoon is able to excite short-wavelength MHD Kelvin-Helmholtz pinch
modes that drive low-amplitude internal shocks along the beam.  These
shocks are not strong however, and it likely that they could only play
a secondary role in the formation of the bright knots observed in
protostellar jets.


\end{abstract}

\clearpage

\section{Introduction}

While magnetic fields seem to play a fundamental role in the production
and initial collimation of astrophysical jets (e.g. K\"onigl \& Ruden
1993, Shu {\it et al.} 1995), they  have been neglected in most of the
analytical and numerical modeling of the structure of protostellar jets
since the inferred estimates of their magnitude ($B \sim
10^{-6}~-~10^{-5}$ G) suggest that they may be not dynamically
important along the flow (e.g., Morse {\it et al.} 1993). However,
after amplification by compression behind the shocks at the jet head,
they may become relevant as they are carried backwards with the shocked
gas to fill the cocoon that envelopes the jet. Moreover, as the
magnetized beam propagates through the surrounding medium (which is
possibly also magnetized), it may be stretched or bent and the
associated compression of the field lines may lead to significant
changes in the beam dynamics and the jet head structure.

A considerable amount of theoretical and numerical work on magnetized,
adiabatic, light jets has been done to study the structure and
evolution of extragalactic jets (e.g. see Birkinshaw 1997 for a
review). Most of that work has focused on the study of the $stability$
properties of the beam against the development of the Kelvin-Helmholtz
(K-H) instability at the contact discontinuity between the jet and the
surrounding medium. Such studies indicate that adiabatic,
supermagnetosonic jets are generally unstable to the K-H pinch and
helical modes  which can successfully explain the formation of
structures such as knots, wiggles and filaments in those jets (e.g.,
Hardee {\it et al.} 1992).  Lately, these  MHD studies have been
extended to $heavy$, $adiabatic$ jets (Todo {\it et al.} 1993, Hardee
\& Clarke 1995, Hardee, Clarke \& Rosen 1997), but the general effects
of ${\bf B}$-fields on the global evolution and morphology of
$radiatively$ $cooling$ jets have not yet been explored in numerical
studies.

In the limit of  zero magnetic field,  numerical simulations of {\it
radiatively cooling}, heavy jets (i.e., denser than their surroundings)
$~-~$ a picture believed to be consistent with protostellar jets $~-~$
have shown that thermal energy losses by the jet system have important
effects on its dynamics (e.g., Blondin, Fryxell \& K\"onigl 1990;
hereafter BFK). This early work was extended to three-dimensional
simulations in a series of papers [Stone \& Norman 1993a, 1993b,
Gouveia Dal Pino \& Benz 1993 (GB93), 1994, Chernin {\it et al.} 1994,
Gouveia Dal Pino, Birkinshaw \& Benz 1996, Gouveia Dal Pino \&
Birkinshaw 1996 (GB96)]. All these studies have revealed that the
cooling jet develops a dense, cold shell of shocked material at the
head which is violently fragmented into clumps by the Rayleigh-Taylor
(R-T) instability.  Moreover,  BFK and GB93 have found that the
development of K-H modes along the beam are inhibited by the presence
of cooling. Recently, Hardee \& Stone (1997) and Stone, Xu \& Hardee
(1997) (see also Massaglia {\it et al.} 1992) have examined the
dynamics of K-H unstable cooling jets. In the case of a
supermagnetosonic jet with longitudinal magnetic field, the linear
analysis of Hardee \& Stone indicates that as long as it does not
dominate the pressure, the magnetic field does not significantly impact
the qualitative differences in K-H stability properties between
adiabatic and cooling jets.

In the present work, we aim to extend these prior studies by examining
the nonlinear effects of magnetic fields (close to equipartition with
the gas) in the structure of radiatively cooling, heavy jets and by
testing their importance on the dynamics of protostellar jets. A
preliminary step in this direction was made in a previous work (Gouveia
Dal Pino \& Cerqueira 1996). Here we mainly concentrate on the effects
of the magnetic field in the jet head structure. In a forthcoming
paper, we will explore the details of the magnetic field effects over
the whole structure of the jet, covering an extensive range of
parameters (Cerqueira \& Gouveia Dal Pino 1998, hereafter CG98).  In \S
2, the numerical method used is briefly described and in \S 3, the
results of the simulations are shown. The conclusions and the possible
implications of our results to protostellar jets are addressed in \S
4.

\section{Numerical Method, Initial and Boundary Conditions}

To simulate the jets in the presence of magnetic fields, we have
employed the smooth particle magnetohydrodynamics (SPMHD) technique
(see, for instance, Stellingwerf \& Peterkin 1990; Meglicki 1995, for
an overview of the method). We solve the MHD equations in the ideal
approximation using a modified version of the three-dimensional SPH
code previously employed by GB93 and GB96 in the investigation of the
evolution of purely hydrodynamical (HD) jets. (A more detailed
discussion with validation tests of the SPMHD will be presented in
CG98.)

Our computational domain is a 3D rectangular box filled with particles
which are initially distributed in a cubic lattice array and which
represent the ambient gas.  The jet of radius $R_j$ is continuously
injected into the bottom of the box and propagates through the ambient
medium up to a distance of $x \approx 30 R_j$.  In the transverse
directions ($y$ and $z$) the box has dimensions $\approx 24 R_j$.  We
use outflow conditions for the box boundaries (e.g., GB96).  The
initial resolution, as defined by the initial smoothing length of the
particle is $0.4 R_j$ and $0.2 R_j$ in the ambient gas and the jet,
respectively. The adiabatic index of the ambient medium and the jet was
assumed to be $\gamma=5/3$, and an ideal equation of state is used. The
radiative cooling, due to collisional excitation and recombination, is
implicitly calculated using a time-independent cooling function for a
gas of cosmic abundances cooling from $T \simeq 10^6$ to $10^4$ K (the
cooling is set to zero for $T ~{\rm <} ~ 10^4$ K; see GB93).

We assume two different initial configurations for the magnetic field.
In one case, we consider an initially  constant longitudinal magnetic
field permeating both the jet and the ambient medium $\vec{B}
=(B_0,0,0)$. This kind of configuration is suggested by observations
that some protostellar jets appear to be aligned with the main
direction of the local interstellar magnetic field (e.g., Appenzeller
1989).  The second geometry adopted here is a force-free helical
magnetic field which also extends to the ambient medium and whose
functional dependence is given by equations 19 to 21 of Todo {\it et
al.} (1993).  In these equations, the maximum strength of the magnetic
field in the system corresponds to the magnitude of the longitudinal
component of the field at the jet axis. The azimuthal component attains
a maximum value ($B_\phi = 0.39B_0$) at $\sim 3 R_j$ (see CG98). The
pitch angle at $1 R_j$ is $\sim 19^{\circ}$.

The parameters of the simulations were chosen to resemble the
conditions found in protostellar jets. In order to compare with
previous purely hydrodynamical simulations (e.g., GB96), we adopt a
number density ratio between the jet and the ambient medium
$\eta=n_j/n_a=3$, $n_j=600$ cm$^{-3}$, an ambient Mach number $M_a =
v_j/c_a= 24$ (where $v_j$ is the jet velocity and $c_a$ is the ambient
sound speed), $v_j \simeq 398$ km s$^{- 1}$, and $R_j = 2 \times
10^{15}$ cm. In the MHD simulations, we assume an initial
$\beta_j=p_{th,j}/p_{B,j} =1$ (the thermal to the magnetic pressure
ratio at the jet axis), which corresponds to a maximum initial value
$B_0 \approx 83\mu$G.  The corresponding initial Alfv\'en and
magnetosonic jet Mach numbers are $M_{A,j} = v_j/v_A \simeq 38$ and
$M_{ms,j} = v_j/(v_{A,j}^2 + c_j^2) \simeq 28$, respectively. (See CG98
for a wider range of parameters.)

\section{Results}

Figure 1 (Plate 1) depicts the density in a midplane section of the
head of three supermagnetosonic, radiatively cooling jets after they
have propagated over a distance $\approx 30 R_j$, at $t/t_d = 1.65$
(where $t_d=R_j/c_a \approx 38$ years).  The top jet is purely
hydrodynamical (HD) ($\beta_j = \infty$, $M_{ms,j} = 42$); the middle
jet has an initial constant longitudinal magnetic field configuration
(with $\beta = 1$, $M_{ms,j} = 28$); and the bottom jet has an initial
helical magnetic field (with axial $\beta = 1$ and $M_{ms,j} = 28$). In
the HD case, the ratio of the radiative cooling length in the
post-shock gas behind the bow shock to the jet radius is
$q_{bs}=d_{cool,bs}/R_j \simeq 8$ and the correspondig ratio behind the
jet shock is $q_{js} \simeq \eta^{-3}q_{bs} \simeq 0.3$ (GB93). These
cooling length parameters imply that within the head of the jet, the
ambient shocked gas is almost adiabatic whereas the shocked jet
material is subject to rapid radiative cooling. Similar initial
conditions are obtained for the MHD cases. The time evolution of the
corresponding velocity distribution is presented in Figure 2.

In the pure HD jet (Figs. 1 and 2, top), we can identify the same basic
features seen in previous work (e.g, GB93). At the head of the jet, the
working surface develops a cold dense shell (with maximum density
$n_{sh}/n_{a} \sim 210$, at $t/t_d = 1.65$) due to the cooling of the
shock-heated jet material. This shell is responsible for most of the
emission of the jet. It becomes R-T unstable (GB93) and breaks into
clumps which spill into the cocoon.  The density of the shell also
undergoes oscillations with time which are caused by global thermal
instabilities of the radiative shocks (see e.g., Gaetz, Edgar \&
Chevalier 1988, GB93).

The cold clumps seen in the HD jet also develop in the shell of the MHD
jet with  longitudinal field (Figs. 1 and 2, middle), but in this case
they detach from the beam as they are expelled backwards to the cocoon.
The fragmentation of the shell is also accompanied by density
oscillations (due to the global thermal instability) but with a maximum
amplitude  which is smaller than that of the HD jet ($n_{sh}/n_a
\approx 155$, at $t/t_d=1.65$).

In the MHD jet with helical magnetic field (Figs. 1 and 2, bottom), a
cold shell also develops initially at the jet head, but the toroidal
component which is initially less intense than the longitudinal
component (by a factor $\leq 2.5$) is amplified by compression in the
shocks at the head (by a factor $\approx 5$). This amplification
reduces the density enhancement and increases the cooling length behind
the jet shock. As a consequence, the shell tends to be stabilized
against the R-T instability and the fragmentation and formation of
clumps is thus inhibited (see also CG98). The dense shell with clumps
seen in the pure HD jet (Fig. 1, top) is replaced in this helical field
case by an elongated, narrow plug of low-density material ($n_p/n_a
\approx 5$) which is pushed by magnetic forces towards the jet axis and
is mixed with the beam material.

Figure 2 also shows the development of some pinching along both MHD
jets. In the pure HD case, constriction occurs only close to the jet
head where the beam is overconfined by the gas pressure of the cocoon.
In the MHD jet with helical field, the toroidal component ($B_{\phi}$)
is amplified by compression in the shocks at the head, and advected
back with the shocked jet material to the cocoon. The associated
magnetic pressure ($\sim B_{\phi}^2/8 \pi$) causes an increase in the
total pressure of the cocoon, relative to the pure HD case. The
increase in the total pressure and the inherent discreteness of the SPH
code in the beam excites (the fastest growing) small-wavelength pinch
modes of the K-H MHD instability (see, e.g., Birkinshaw 1997, Cohn
1983). These modes overconfine the beam and drive the approximately
equally spaced internal shocks seen in the MHD jet with helical field
at the positions x $\simeq -1R_j$, 3$R_j$ and 7$R_j$ (Fig. 2, bottom).
The fact that both, the gas pressure and the magnetic pressure in the
cocoon have approximately the same magnitude, and the toroidal magnetic
field exceeds the longitudinal component, suggests that the unstable
modes are driven by both magnetic field effects and the velocity
discontinuity at the interface between the jet and the cocoon.
It is interesting to note that the numerical study of Todo {\it et al.}
(1993) employed a similar helical field geometry. In that case however,
the collision of the jet with the ambient cloud led to an MHD kink
instability that is not seen in our simulations.

Along the MHD jet with longitudinal field (Fig. 2, middle), the
increase in the total confining pressure of the cocoon also drives the
development of  K-H MHD instabilities which produce the beam pinching
and  internal shocks seen in the figure. Consistent with the linear
theory for K-H modes in the supermagnetosonic regime, they begin to
appear at a distance $\approx M_{ms,j} R_j$. As expected, this
destabilization length is smaller than that for the pure HD jet.  Also,
we found a close correlation between the pinching zones and the
appearance of more intense reversed magnetic fields at the contact
discontinuity between the jet and the cocoon.  The originally uniform
longitudinal fields are reoriented in the non-parallel shocks at the
head and swept-back into the cocoon. As a consequence, a predominantly
toroidal current density distribution develops around the jet radius
($\vec{J} \simeq \vec{J}_{\phi}$).  Such configuration creates a
$\vec{J} \times \vec{B}$ force ($\sim -J_\phi B_{\parallel}$)
that constricts the beam (this is analyzed in detail in CG98).

\section{Discussion and Conclusions}

We have presented the results of fully three-dimensional simulations of
supermagnetosonic, overdense, radiatively cooling jets with two
different initial magnetic field configurations (with magnitude in
close equipartition with the gas).  Our results indicate that magnetic
fields can have significant effects on the morphology and dynamics of
radiatively cooling jets, particularly at the $head$ structure.

As in previous adiabatic MHD calculations (Todo {\it et al.} 1993;
Frank {\it et al.} 1997), the helical field in the cooling jet is
amplified and reoriented by the shocks at the head. This reduces the
post-shock compressibility and thus increases the post-shock cooling
length relative to the pure hydrodynamical (HD) jet. This stabilizes
the shell at the head against the R-T instability. As a result, the
clumps that develop by fragmentation of the dense shell at the head of
the HD jet do not appear in the helical magnetic jet.  Instead, the
clumpy, dense shell is replaced by an elongated plug of low-density
material which is pushed to the jet axis by the development of $\vec{J}
\times \vec{B}$ forces. This structure resembles the nose cone seen in
previous adiabatic simulations of strongly magnetized jets with
toroidal fields (e.g., Clarke, Norman \& Burns 1986, K\"ossl, M\"uler
\& Hillebrandt 1990).  On the other hand, a cooling jet embedded in a
longitudinal magnetic field retains the clumps seen at the head of the
HD jet. The fact that the fragmented, clumpy shell resembles the knotty
structure often observed in Herbig-Haro objects behind the bow shocks
of protostellar jets (e.g., HH1, HH2, HH19, HH12, HH95, HH45, HH47D
$~-~$ Herbig \& Jones 1981; Brugel {\it et al.} 1985; Reipurth 1989;
Heathcote {\it et al.} 1996) suggests that a strong helical magnetic
field geometry is unlikely in these jets. If the jet is permeated by
strong magnetic fields, these are probably predominantly longitudinal,
at least in the outer regions of the jet far from the driving source.

Over the simulated time and length scales, in both initial magnetic
field configurations but not in the HD run, approximately equally
spaced internal shock are produced by K-H MHD reflection modes. These
shocks propagate forward with velocities similar to that of the head of
the jet ($v_{bs} \approx 250$ km s$^{-1}$). The mean distance between
the internal shocks $\approx 3R_j$ is in agreement with the observed
jet knots. However, the weakness of these shocks ($n_{is}/n_a \lesssim
10 $) makes it doubtful that they could produce by themselves knots as
bright as those observed in protostellar jets.

Finally we note that the beam pinching seen in the MHD cooling jets
discussed here is also found in adiabatic jets with similar initial
conditions (see CG98). Their strength is, however, larger in the
adiabatic case. This result is consistent with previous numerical
analyses comparing pure HD jets with and without cooling (e.g., BFK,
GB93), which have indicated that the presence of radiative cooling
tends to reduce the strength and the number of internal shocks in the
jets. This result is also compatible with the predictions of the linear
stability theory  (Hardee \& Stone 1997) when applied in the context of
the cooling function employed in the present work.

\acknowledgements
The work of A.H.C and E.M.G.D.P. was partially supported by the
Brazilian agencies FAPESP and CNPq. The work of M.H. was supported in
part by a Director funded Postdoctoral Fellowship at the Los Alamos
National Laboratory. We are also thankful to an anonymous referee for
his remarks. The simulations were performed on a DEC 3000/600 AXP
workstation, whose purchase was made possible by FAPESP.

\newpage

\centerline {\bf Figure Captions}

\noindent {\bf Figure 1:} Color-scale representation of the midplane
density of the head of a hydrodynamical jet (top); an MHD jet with
initial longitudinal magnetic field configuration (middle); and an MHD
jet with initial helical magnetic field configuration (bottom), at a
time $t/t_d=1.65$ ($t_d=R_j/c_a \simeq 38$ years). The initial
conditions are: $\eta=n_j/n_a=3$, $n_a=200$ cm$^{-3}$, $M_a=24$, $v_j
\simeq 398$ km s$^{-1}$, $q_{bs}\simeq 8$ and $q_{js} \simeq 0.3$. The
initial $\beta = 8\pi p/B^2$ for the MHD cases is $\beta \simeq 1$.
The color scale (from minimum to maximum) is given by:  black, grey,
white, orange, yellow and blue.  The maximum density reached by the
shell at the head of the jets is $n_{sh}/n_a \simeq 210$ (top),
$n_{sh}/n_a \simeq 153$ (middle), and $n_{sh}/n_a \simeq  160$
(bottom).

\noindent {\bf Figure 2:} Midplane velocity field distribution
evolution of the head region of the jets of Fig. 1: hydrodynamical jet
(top); MHD jet with longitudinal magnetic field (middle); and MHD jet
with helical magnetic field (bottom). The initial parameters are the
same as in Fig. 1. The times and the jet head positions are:
$t/t_d=1.40$ and x $\simeq 25R_j$ (left); $t/t_d=1.60$ and x $\simeq
29R_j$ (middle); $t/t_d=1.65$ and x $\simeq 30R_j$ (right).

\end{document}